# Higher-Order Modes Are More Robust: The Origin of Bending Insensitivity in Multimode Fibers and Its Exploitation for Imaging

**XIANGQIAN CHEN,[1,2,3] LONG HE,[1,2,3] QIANKAI XIA,[1,2,3] JIALIANG LV,[1,2,3] YUXUAN LUAN,[1,2,3] CHEN YANG,[1,2,3] LIXIN XU,[1,2,3] PEIJUN YAO,[1,2,3,*]**

*1 Department of Optics and Optical Engineering, University of Science and Technology of China, Hefei, Anhui 230026, China*

*2 State Key Laboratory of Opto-Electronic Information Acquisition and Protection Technology, Anhui University, Anhui, Hefei, China*

*3 Advanced Laser Technology Laboratory of Anhui Province, Hefei, Anhui 230026, China*

**yap@ustc.edu.cn*

**abstract：** Multimode fibers (MMFs) hold great promise for minimally invasive imaging, yet their extreme sensitivity to bending perturbations severely hinders practical applications. Starting from mode coupling theory, we theoretically prove and experimentally verify that higher-order modes exhibit significantly greater bending robustness than lower-order modes. We reveal that bending-induced changes in the output speckle arise predominantly from alterations in the mode effective propagation constants, whereas the modal power redistribution caused by inter-modal coupling is negligible. We propose a physics-inspired dual-encoder neural network. The network separately extracts features from a reference speckle pattern that corresponds to a circular field input and an information-carrying speckle pattern acquired under the same fiber state, then employs a differential fusion module to decouple the image information from environmental perturbations, and subsequently recovers the image after excluding the perturbation. Our method significantly enhances the anti-bending capability of MMF imaging systems, offering a new paradigm that integrates physical insights with deep learning for robust fiber-optic imaging.

## 1. Introduction

Multimode fibers (MMFs), owing to their exceptionally high information capacity and ultrathin, flexible geometry, are regarded as an ideal medium for next-generation ultra-thin endoscopes[1,2]. At the laboratory research level, various methods have been developed to enable image transmission and reconstruction through MMFs, including the transmission matrix method[3,4,5], digital phase conjugation[6,7], compressive sensing[8,9,10], and deep learning[11,12], which have preliminarily demonstrated basic imaging capabilities. However, most of these approaches rely on the assumption that the fiber state remains unchanged. Once the fiber state is perturbed, the output quality of the established model degrades significantly, and recalibration is often time-consuming. Consequently, the stability and real-time performance of imaging systems have become key bottlenecks hindering the practical application of multimode fiber endoscopes.

T. Čižmár et al. established an accurate theoretical model and, for the first time, realized the computational prediction of the transmission matrix of bent multimode fibers[5]. Experiments demonstrated that light transmission through the fiber remains highly predictable even under large-curvature bending, based on which they successfully demonstrated imaging through a bent fiber without the need for experimental calibration. S. Li et al. proposed a compressive-sensing-based method for reconstructing the transmission matrix of multimode fibers[13]. By exploiting the sparsity of the propagation-invariant mode basis, high-fidelity reconstruction can be achieved with only about 5% of the measurements required by conventional methods, and the compression ratio for imaging can be as low as 1%. Both approaches improve, to a certain extent, the efficiency of transmission matrix reconstruction when the fiber state changes. However, the transmission matrix method itself is still limited by the requirement for high phase-detection accuracy and the dependence on a stable state of the multimode fiber, making it difficult to apply in practical scenarios where the fiber bending state changes rapidly.

S. Yan et al. utilized deep learning techniques, training a neural network on a large number of random bending configurations so that the model can reconstruct the input image solely from the output speckle intensity without knowledge of the specific fiber configuration[14], offering a new data-driven approach for fiber imaging in dynamic environments. Building on this, Fan et al. further introduced a continual learning mechanism and proposed the DI-GAN framework[15], which enables the model to sequentially learn different bending states under extreme memory constraints and prevent forgetting by synthesizing data from itself, ultimately achieving adaptive focusing across states using only reflected speckle, thereby significantly improving the efficiency of dynamic calibration and the focusing quality. Yang et al. proposed the STABLE (spatial-frequency tracking adaptive beacon light-field-encoded) dynamic imaging framework[16], which shifts the monitoring target from the distal speckle image to a spatial frequency domain beacon and exploits the deterministic mapping between beacon shift and fiber deformation to achieve sub-millisecond tracking of bending and twisting. This strategy fundamentally circumvents the "generalization" challenge, realizing stable imaging under dynamic deformation without the need for distal access, and improves the resolution to approximately $\lambda/3NA$. For the first time, high-resolution imaging capability was validated in the gastrointestinal tract of living small animals, opening a new technical pathway for flexible endoscopy toward clinical dynamic applications. Nevertheless, the STABLE framework is currently still limited by system synchronization bottlenecks, and the overall imaging frame rate has not yet met clinical real-time requirements; future breakthroughs will require hardware acceleration strategies such as FPGA.

To address the above bottlenecks, other studies have explored new technical pathways from different perspectives. Liu et al. proposed an all-fiber high-speed image detection method that encodes spatial information into temporal waveforms by exploiting the large intermodal dispersion in multimode fibers[17], achieving a frame rate of 15.4 Mbps, a shutter time of 45.1 ps, and a single-shot recording depth of 10,000 frames using only a single ultrafast photodiode. This work overcomes the speed limitation imposed by conventional cameras on fiber imaging and provides a new all-fiber, highly integrated solution for observing transient biomedical phenomena. Zheyu Wu et al. replaced conventional multimode fibers with ring-core fibers[18], utilizing the weakly coupled mode group structure formed by orbital angular momentum modes to effectively suppress the disturbance of bending and vibration on the output speckle pattern, thereby enhancing the stability of the imaging system at the fiber-device level and reducing system complexity. Building on this, they further proposed a differential neural network method based on dual-polarization speckle[19], which extracts and maps the perturbation features in two orthogonal polarization directions and eliminates them through differentiation, significantly enhancing the system's robustness under strong perturbation environments. Recently, Zheyu Wu et al. shifted their research focus to graded-index multimode fibers[20]; however, under dynamic perturbations such as vibration and bending, they could only transmit and detect image information streams in the form of pattern recognition, and have not yet been able to directly recover complex image information.

Although the output speckle pattern of a multimode fiber evolves dramatically as the bending state changes, existing studies have shown that certain components within it vary more slowly. Loterie et al. designed an S-shaped bending device that can slide along the fiber axis and found that, as the bending position moves, the transmission properties of the fiber can remain stable within a certain range, and that the peripheral region of the core exhibits significantly stronger bending resistance than the central region, indicating that the higher-order mode distribution region has a higher tolerance to morphological changes[21]. Li et al. approached the problem from the perspective of the memory effect, revealing a "quasi-radial" memory effect in multimode fibers, and, with the aid of guide-star-assisted imaging, achieved scanning imaging over a certain range under bending; their study showed that the scanning range is closely related to the focal position of the guide star within the core, and that the farther the focal position deviates from the center, the larger the achievable scanning range[22]. Xu et al. further discovered in their research on perturbation-resistant imaging that under continuous deformation, the peripheral region of the speckle pattern maintains strong correlation while the central region becomes nearly completely randomized; based on this observation, they used a deep learning network to extract invariant fiber features from the peripheral region, achieving high-fidelity image reconstruction and high-precision object recognition[23]. Together, these works demonstrate that the optical field in the peripheral region of a multimode fiber (corresponding to the higher-order mode distribution region) possesses significantly stronger robustness against bending perturbations, providing an important physical foundation for bending-resistant imaging.

In this paper, from the perspective of mode coupling theory, the physical mechanism behind the "rapid change in the center and gradual change in the periphery" observed in the output speckle of multimode fibers, offering theoretical support for utilizing the peripheral speckle region to achieve bending-resistant imaging. Accordingly, we propose a physics-inspired dual-branch encoder neural network. This network takes the speckle produced by a circular optical field input under the same fiber state as a reference, performs joint feature extraction with the measurement speckle carrying unknown object information, and then eliminates environmental perturbations through a differential fusion module, thereby achieving reliable recovery of the original image information. Meanwhile, to ensure that the peripheral speckle region contains complete image information, a spatial information mixing module is designed at the fiber input end, and the peripheral region of the speckle pattern is fixedly selected as the network input. This approach significantly enhances the bending resistance of the multimode fiber imaging system while reducing computational complexity, providing a new paradigm for robust fiber imaging that integrates physical insights with deep learning.

## 2. Principle

### *A. Mode Coupling Theory under Bending Perturbations*

The output image of a multimode fiber can be regarded as the coherent superposition of the eigenmodes transmitted within the fiber on the receiving screen. The coherent superposition pattern is influenced by intermodal dispersion and the power distribution among modes. The optical field transmitted in the fiber is expressed as:

$$\Psi\,(r,\varphi,z)=\sum\nolimits_{l}A_l(z)\Psi_l(r,\varphi)\exp(-j\beta_l z) \tag{1}$$

where $\Psi_l(r,\varphi)$ represents the optical field distribution of the different eigenmodes of the unbent fiber, $A_l(z)\exp(-j\beta_l z)$ represents the normalized complex amplitude of the different modes, and $\beta_l$ is the longitudinal propagation constant of the different modes. When considering the influence of bending on the output speckle, it is necessary to study the effect of multimode fiber bending on the internal mode coupling. The diameter of multimode fibers is mostly 125 μm. Under normal circumstances, the bending radius R of the fiber is much larger than the fiber diameter d, (R≫d). Therefore, the perturbation method is often employed to investigate the effect of fiber bending on the mode coupling within the fiber[24,25,26]. The coupling behavior between modes can be expressed as:

$$\frac{dA_l}{dz}=\sum\nolimits_{s}k_{ls}A_s\exp[i(\beta_l-\beta_s)z] \tag{2}$$

$$\kappa_{ls} = -\mathrm{j}\frac{k_0^2}{2\beta_l}\frac{\iint \mathbf{E}_l^* \Delta(n^2)\mathbf{E}_s \mathrm{d}S}{(\iint |E_l|^2 \mathrm{d}S \iint |E_s|^2 \mathrm{d}S)^{\frac{1}{2}}} \tag{3}$$

Where $\kappa_{ls}$ is the mode coupling coefficient, $E$ denotes the electric field distribution of the different modes, and $\Delta(n^2)$ represents the perturbation induced by fiber bending.

We analyze the explicit expression for $\Delta n^2$ by considering both the bending-induced waveguide structural change and the photoelastic effect. [27,28,29,30]. The effect of bending on the fiber can be expressed as:

$$\Delta n^2 = 1.56\frac{x}{R}{n_1}^2 \tag{4}$$

The detailed calculation process can be found in Supplementary Material 1.

Using Equation (2) and (3), we adopt an iterative method to solve for the amplitude variation of different modes after propagating a distance $(z)$. The second-order iterative result is as follows, with the detailed solution process provided in Supplementary Material 2:

$$A_l(z)^{(2)} = A_l(0)\left(\sum_{s\neq l}\frac{k_{ls}}{i(\beta_l-\beta_s)}A_s(0)\{exp[i(\beta_l-\beta_s)z-1]\} * e^{i\delta\beta_l z} + 1\right)e^{-i\delta\beta_l z} \tag{5}$$

where $\delta\beta_l = -\sum_{s\neq l}\frac{k_{ls}k_{sl}}{(\beta_l-\beta_s)}$. It can be noted that in the second-order approximation, bending not only causes fluctuations in the power distribution among modes but also introduces an additional phase term related to the propagation distance, which can be approximately regarded as a change in the propagation constant of the modes when the fiber is bent[25].

### *B. Numerical Simulation and Experimental Verification*

To further analyze the specific influence of changes in the fiber bending state on the output image of a multimode fiber, we set up an initial experimental setup, as shown in Fig. 1. The laser wavelength is 639 nm, and the multimode fiber has a core/cladding diameter of 50/125 μm. After passing through a beam expansion system, the laser illuminates a digital micromirror device (DMD), so that the reflected optical field carries the image information loaded on the DMD. The light is then coupled into the multimode fiber via a focusing lens, and the speckle pattern output from the multimode fiber is captured by a CMOS camera. The output end of the fiber and the CMOS camera are placed together on a translation stage. As the translation stage moves, the bending state of the multimode fiber changes; the maximum travel range of the translation stage is 4 cm. Based on this experimental setup, we determined the relevant parameters for the simulation calculations (see Supplementary Material 3 for details of parameter selection), and we actually collected a series of output speckle patterns from the multimode fiber as the bending state changed, in order to facilitate comparison with the simulation results.

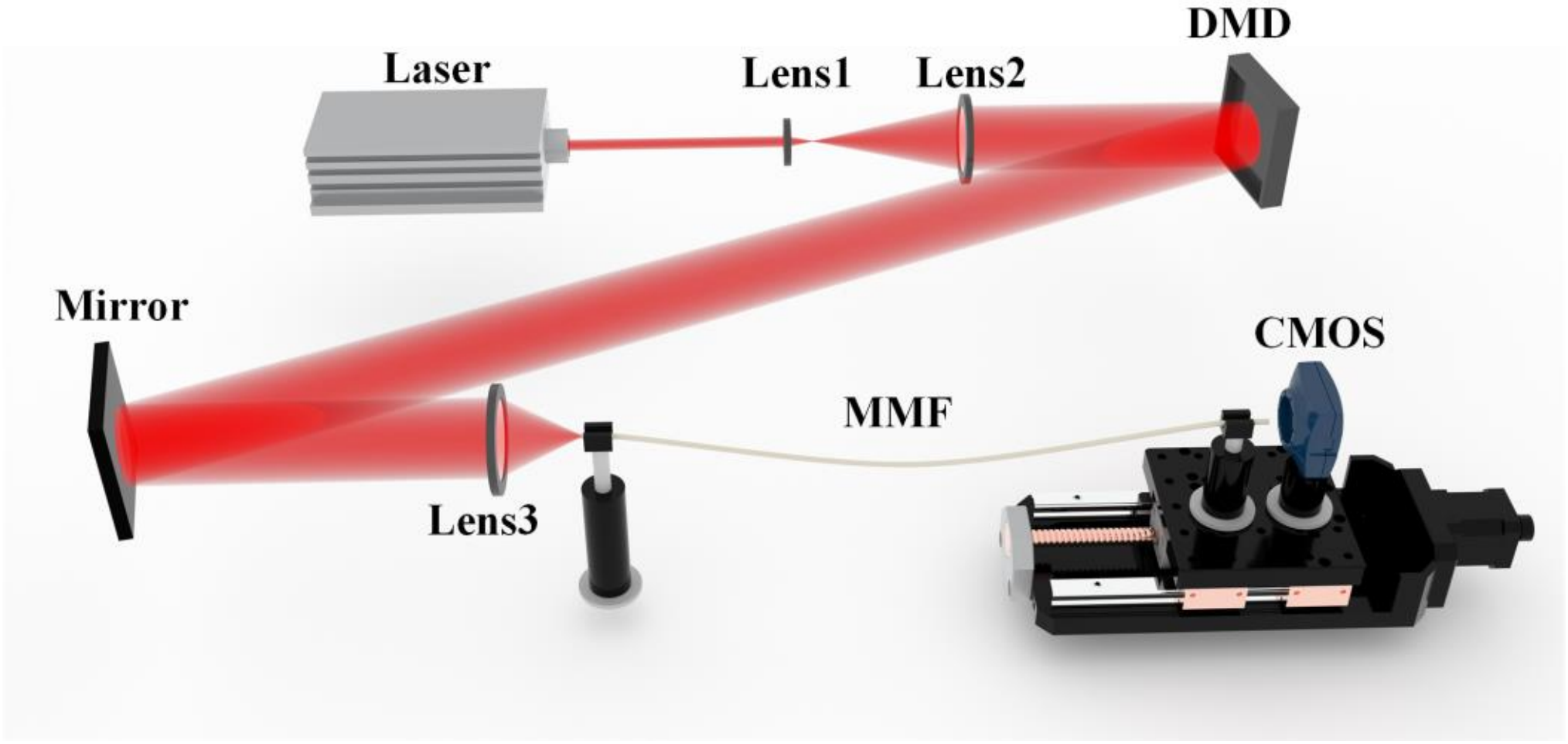

Figure 1 Experimental setup for studying the influence of fiber bending on the output image of a multimode fiber.

In the calculations, the weakly guiding approximation was used for the multimode fiber, and the linearly polarized (LP) modes ($LP_{mn}$) were taken as the eigenmodes of the fiber to represent the optical field transmitted in the fiber. When calculating the mode coupling coefficients, only coupling between modes of the same polarization state was considered, because pure bending of an ideal fiber does not cause polarization coupling between its eigenmodes. Admittedly, these two approximations are obviously too idealized and cannot perfectly replicate the actual mode coupling results inside the fiber. However, we can still use them to obtain some characteristics of the mode coupling inside the multimode fiber under bending, which can then be verified experimentally.

We first collected the output images of the multimode fiber as the fiber bending state changed, with the DMD loaded with a fully bright circular pattern. Under this condition, the exposure range of the loaded pattern on the DMD already covered the maximum acceptance aperture of the multimode fiber. Images were captured at every 1 mm movement of the translation stage, yielding a total of 40 images. Figure 2(a) shows the images acquired at every 4 mm of stage movement, i.e., 10 images in total. Meanwhile, we also calculated the variation of the multimode fiber output image with the bending state under the same condition, as shown in Fig. 2(c). A direct observation of the two image sequences in Fig. 2(a) and Fig. 2(c) clearly reveals that in both the experimental and simulated output images, the central region changes drastically while the peripheral region changes slowly. To visualize this more intuitively, we divided the experimentally and numerically obtained output images into two parts: a central circular region and a surrounding annular region, and calculated the Pearson correlation coefficient (PCC) between the output image at each bending state and the initial-state image, as shown in Fig. 2(f) and Fig. 2(g). It can be seen that, for both the experimentally acquired and simulated results, as the fiber bending state changes, the PCC of the central region of the multimode fiber output image decreases much faster than that of the annular region farther from the center. This provides a preliminary indication that the influence of fiber bending on the output speckle is mainly manifested in the central region of the image, while its impact on the peripheral annular region is relatively weak.

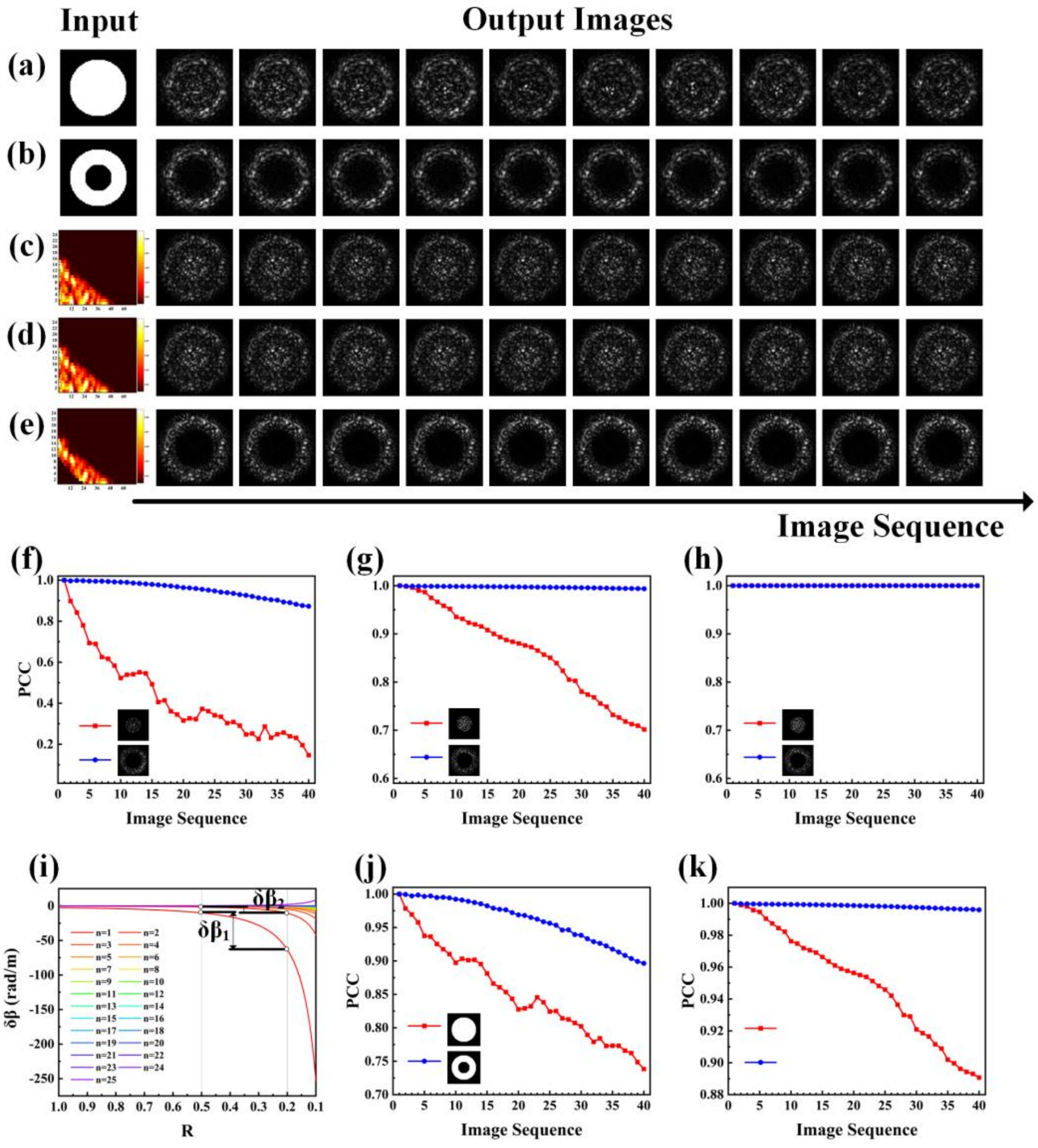

Figure 2 (a) Output images of the multimode fiber acquired every 4 mm of translation stage movement when a circular pattern is loaded on the DMD; (b) output images acquired every 4 mm of translation stage movement when an annular pattern is loaded on the DMD; (c)–(e) simulated output images: (c) all modes present, considering both amplitude and phase variations;

(d) all modes present, considering only amplitude variations; (e) only higher-order modes present, considering both amplitude and phase variations. The heatmaps in (c)–(e) show the mode amplitude coefficients of different $LP_{mn}$ modes, with the horizontal axis representing *m* and the vertical axis representing *n*; (f) Pearson correlation coefficient (PCC) curves of the experimentally acquired images corresponding to (a): red for the central circular region, blue for the peripheral annular region; (g) PCC curves of the simulated images considering both amplitude and phase variations: red for the central region, blue for the annular region; (h) PCC curves of the simulated images considering only amplitude variations: red for the central region, blue for the annular region; (i) change in the effective propagation constant $\delta\beta$ as a function of bending radius *R* for fiber modes $LP_{mn}$ ($m = 0$, $n = 1$–$25$); (j) PCC curves of the experimentally acquired images: red for the circular pattern loaded on the DMD, blue for the annular pattern loaded on the DMD; (k) PCC curves of the simulated images: red for the case where both high- and low-order modes are present, blue for the case where only higher-order modes are present.

To further analyze the causes of the variation in the multimode fiber output image, we return to expression (5). The influence of changes in the fiber bending state is mainly manifested in two aspects: the power distribution among different modes and the variation in the effective propagation constants of different modes. First, we set $\delta\beta_l = 0$ and considered only the result of energy exchange among modes. The simulated results are shown in Fig. 2(d). We also calculated the PCC trends for the central region and the peripheral annular region, as shown in Fig. 2(h). It can be seen that when only intermodal energy exchange is considered, the output image remains essentially unchanged and does not exhibit the behavior observed in the experiment — namely, drastic changes in the central region and slow changes in the peripheral annular region. This indicates that the variation of the multimode fiber output image has little correlation with the power fluctuations among modes. Instead, the variation characteristics of the output image in the experiment are very directly related to the change in the effective propagation constant of the modes induced by bending.

To further explain the correlation between the change in the effective propagation constant of the modes and the variation characteristics of the output speckle of the multimode fiber, we calculated the relationship between $\delta\beta_l$ and the fiber bending radius *R* for the $LP_{mn}$ modes with $m = 0$, $n = 1$–$25$, as shown in Fig. 2(i). It can be seen that under the same change in bending radius, the variations of $\delta\beta_l$ for different modes are not identical. This leads to a continuous change in the relative phase differences among different modes at the output facet as the bending radius of the multimode fiber varies. Since the image captured by the camera is the coherent superposition of the propagating optical fields of different modes on the camera receiving plane, the captured image also changes drastically with the fiber bending state. Comparing $\delta\beta_1$ and $\delta\beta_2$ for the modes with $n = 1$ and $n = 2$ when the bending radius changes from 0.5 m to 0.2 m, it is clearly seen that $\delta\beta_1$ exhibits a larger variation amplitude.

By comparing the changes in $\delta\beta_l$ for all modes, it can be found that within a certain range, the smaller $n$ is, the larger the corresponding change in the effective propagation constant. Although there is a slight increase in the variation amplitude of the effective propagation constant for larger $n$, the magnitude of change is still significantly smaller than that for smaller $n$. We refer to such modes that are far from the fiber cutoff condition and have drastic changes in the effective propagation constant as low-order modes (here, "low-order modes" is a general definition for this fiber specification, and there is no strict boundary in terms of $m$ and $n$), while the remaining modes that are close to the cutoff condition and exhibit slow changes in the effective propagation constant are referred to as high-order modes.

The image captured by the camera is the intensity distribution formed by the coherent superposition of the optical fields of different modes after propagating to the camera receiving plane. In a multimode fiber, low-order modes generally have smaller divergence angles compared to high-order modes and primarily contribute to the central region of the image on the camera receiving plane, whereas the contribution of high-order modes is mainly concentrated in the peripheral annular region of the image. Since the effective propagation constants of low-order modes undergo the most drastic changes as the bending state of the multimode fiber varies, the speckle pattern observed at the camera receiving plane exhibits strong variation in the central region and weak variation in the periphery.

To further verify this conclusion, we adjusted the pattern loaded on the DMD to an annulus. Due to the conservation of the K-vector during light transmission in the multimode fiber, the intensity distribution of the light field received by the camera also exhibits an annular shape, indicating that the modes predominantly excited in the fiber at this point are high-order modes. Under this condition, we again varied the bending state of the multimode fiber and acquired images, as shown in Fig. 2(b). We also calculated the PCC trend of the acquired images relative to the initial-state image, and compared it with the PCC trend for the circular pattern, as shown in Fig. 2(j). Similarly, we simulated the variation of the multimode fiber output image when low-order mode energy is absent, as shown in Fig. 2(e), and calculated its PCC trend, then compared it with the case where both high-order and low-order mode energies are present, as shown in Fig. 2(k). Both the experimental and simulated results show that when low-order mode energy is nearly absent, all the multimode fiber output images received by the camera exhibit an annular distribution, and the influence of the bending radius variation is significantly weakened. This demonstrates that in multimode fibers, high-order modes possess more stable propagation characteristics compared to low-order modes. In applications such as spatial division multiplexing for information transmission and endoscopic imaging, high-order modes in multimode fibers offer inherent advantages.

It is worth noting that the speckle patterns obtained from simulations and experiments are not entirely identical. We attribute this discrepancy primarily to the difficulty of uniformly exciting the higher-order modes in the multimode fiber through such direct optical field coupling in the experiment. In the simulation, although the mode excitation is randomly sampled from the actual field distribution, the resulting modal distribution is still more uniform. Nevertheless, this difference does not affect the experimental conclusions of this work.

## 3. Result

### *A. Experimental Setup*

We have demonstrated that high-order modes in multimode fibers exhibit greater bending robustness in information transmission applications. To address the susceptibility of multimode fiber-based information transmission to bending perturbations, we constructed an experimental setup for bend-resistant image transmission using high-order modes of a multimode fiber, as shown in Fig. 3(a). The setup is essentially the same as that in Fig. 1. However, due to the K-vector conservation property of multimode fibers, the image information at the center of the DMD is difficult to couple into the region where the high-order modes reside. As a result, the high-order modes of the multimode fiber carry only the information of the peripheral annular region of the image, rather than the complete information of the entire image — a situation that obviously must be avoided when using multimode fibers for image information transmission. Therefore, we incorporated a mode mixing module into the setup of Fig. 3(a). This mode mixing module is made of a very short piece of multimode fiber of the same specification, and the end faces where it connects to the transmission multimode fiber segment are both roughened. Before the image information is coupled into the transmission multimode fiber, we first perform mode mixing on the coupled-in optical field. In this way, the information at the interface between the two fiber end faces is fully mixed between high-order and low-order modes, so that regardless of the spatial light field coupled into the fiber, both high-order and low-order modes in the transmission fiber can be well excited.

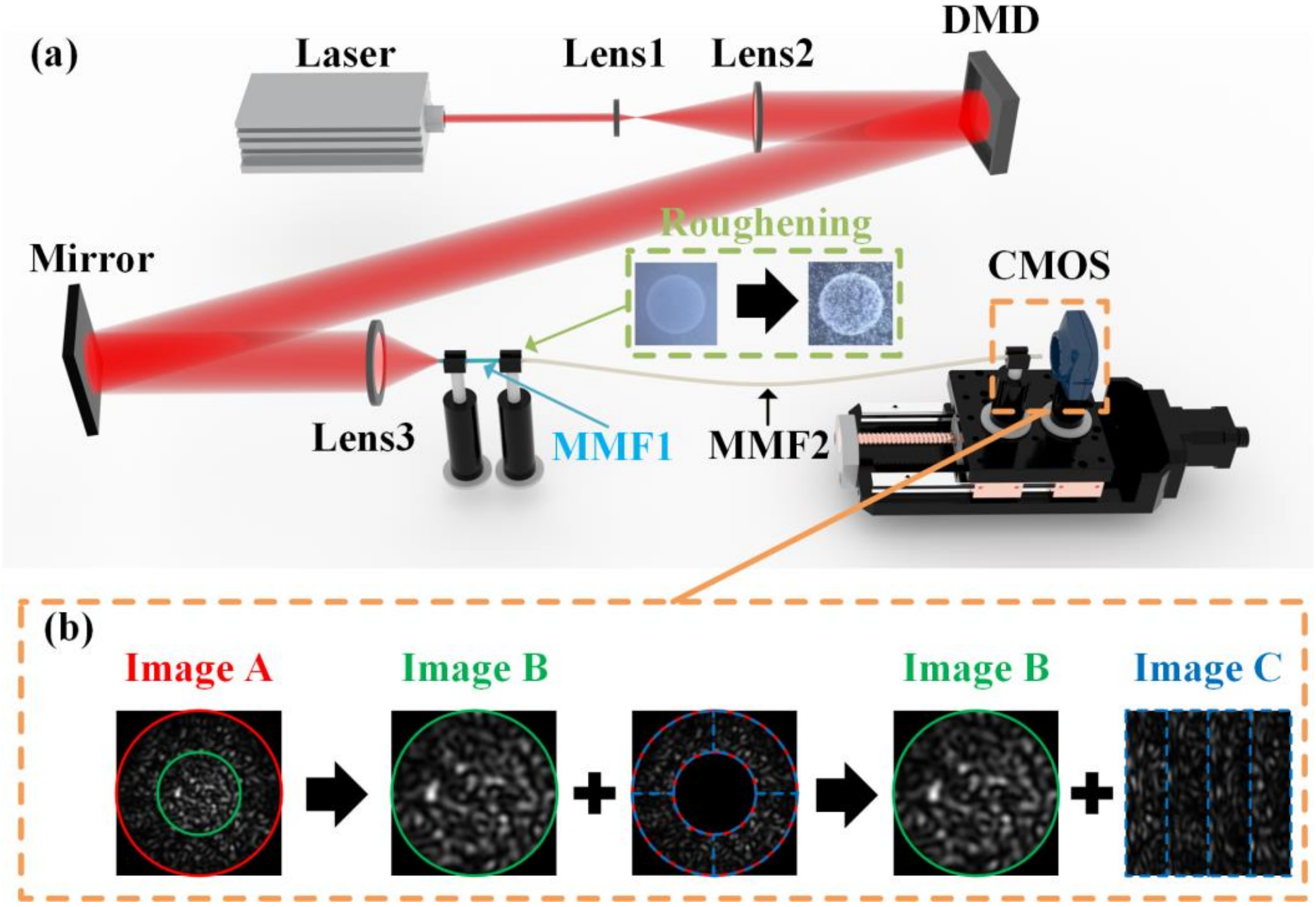

Figure 3 (a) Experimental setup for image information transmission incorporating a mode mixing module. Here, MMF1 is the mode mixing fiber, and MMF2 is the information transmission fiber; both have a core/cladding diameter of 50/125 μm, and their connecting end faces are roughened. (b) After mode mixing, the speckle pattern output from the multimode fiber can be processed through a specific segmentation and recombination method to separately reconstruct the complete information (Image A), the low-order mode information (Image B), and the high-order mode information (Image C).

### *B. Information Integrity of the Separated Mode Components*

To verify whether the mode mixing module can sufficiently mix the image information carried by high-order and low-order modes, we conducted image information transmission experiments while keeping the fiber bending state unchanged. In the experiment, binarized MNIST handwritten digit images were sequentially loaded onto the DMD. For each loaded image, the corresponding output image from the multimode fiber was captured by the camera. The captured images were processed according to the method illustrated in Fig. 3(b) and then paired with the input images to construct three types of image-label pairs: complete information (Image A), low-order mode information (Image B), and high-order mode information (Image C). The first 10,000 images from the MNIST dataset were used, with the first 8,000 image-label pairs serving as the training set and the remaining 2,000 as the test set. In the following, Image A, Image B, and Image C refer to these three different ways of dividing the image information, respectively.

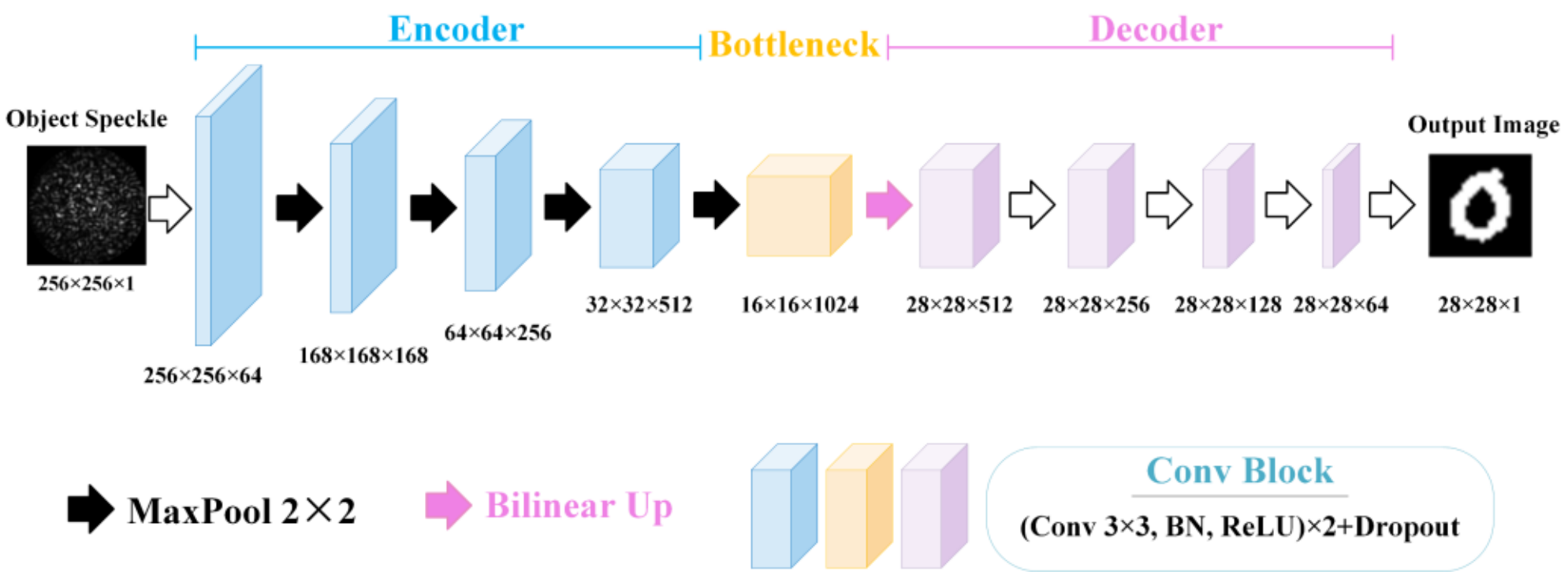

Figure 4 Architecture of the encoder–decoder convolutional neural network (ED-CNN) for image reconstruction from speckle patterns. The network takes a 256×256×1 speckle image as input and outputs a 28×28 reconstructed image. The encoder consists of four cascaded blocks, each comprising convolutional layers (3×3 kernels), batch normalization, LeakyReLU activation, max-pooling, and dropout, progressively reducing spatial resolution while increasing channel depth. A bottleneck layer with additional convolutional layers further extracts features. The decoder upsamples the feature maps from 16×16 to 28×28 via bilinear interpolation, followed by multiple convolutional layers for refinement, and finally applies a 1×1 convolution with sigmoid activation to produce the output.

We implemented a convolutional neural network with an encoder–decoder architecture using TensorFlow to recover 28×28 digit images from 256×256 speckle patterns. The encoder consists of four cascaded encoder blocks, each comprising two convolutional layers (3×3 kernels), batch normalization, LeakyReLU activation (negative slope 0.2), max-pooling (stride 2), and dropout layers to mitigate overfitting. This progressively downsamples the spatial dimensions from 256×256 to 16×16, while the number of channels increases from 64 to 512. A bottleneck layer with two convolutional layers of 1024 channels further extracts deep features. The decoder first upsamples the feature maps from 16×16 to 28×28 via bilinear interpolation to avoid checkerboard artifacts, then passes through multiple convolutional layers that gradually reduce the number of channels from 512 to 64 for refined reconstruction, and finally produces a single-channel grayscale image through a 1×1 convolution with sigmoid activation.

During training, the loss function is a weighted combination of binary cross-entropy and Dice loss (each with a weight of 0.5) to balance pixel-wise classification accuracy and region overlap. The Adam optimizer is employed with an initial learning rate of $1\times10^{-4}$ and gradient clipping (threshold 1.0) to stabilise training. To prevent overfitting and improve generalisation, we incorporate a learning rate reduction strategy (halving the learning rate if the validation loss does not decrease for 3 consecutive epochs, down to a minimum of $1\times10^{-6}$), early stopping with a patience of 10 epochs, and a model checkpoint that saves the best weights based on validation loss. The model is compiled with mean squared error as an auxiliary monitoring metric. Overall, the design achieves a balanced trade-off between reconstruction accuracy and computational efficiency.

For the three training sets—Image A, Image B, and Image C—the same network architecture was used for training to ensure the comparability of the results.

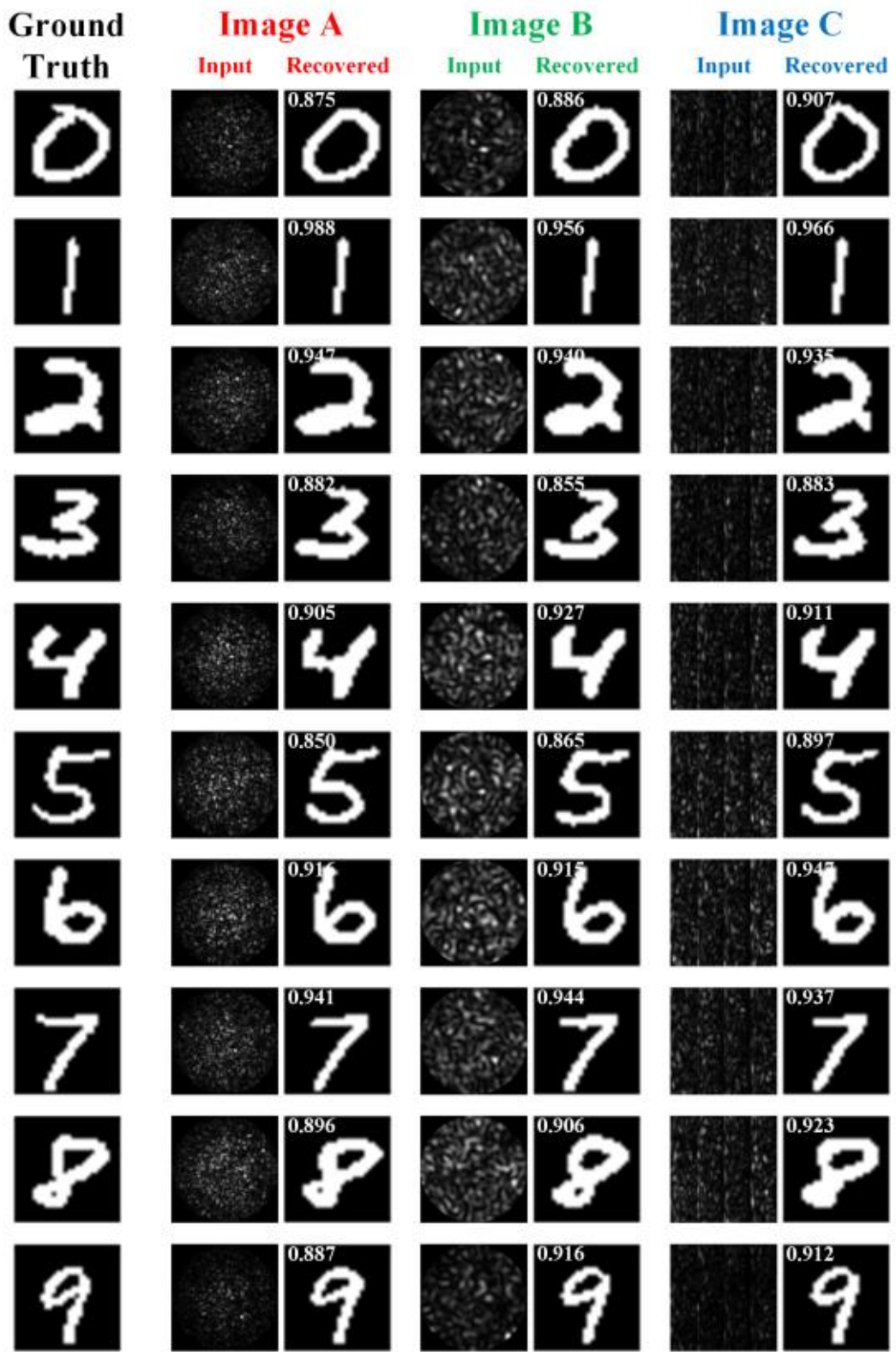

Figure 5. Comparison of recovery results for some images from the test set. The number in the upper left corner of each recovered image indicates its Dice similarity coefficient (DSC) with the ground truth image.

After training, the test sets were respectively input into the trained neural network to obtain the output images, which were then compared with the ground truth images. Since both are binary images, the image recovery task is more akin to an image segmentation task. Therefore, the Dice Similarity Coefficient (DSC) was chosen to evaluate the image recovery quality. The expression of DSC is given by Equation (6), where $A$ and $B$ are both binary images:

$$DSC = \frac{2|A \cap B|}{|A| + |B|} \tag{6}$$

The experimental results show that the average DSC for the test set corresponding to Image A is 0.877, for Image B is 0.871, and for Image C is 0.887. The recovery results for some images from the three test sets can be seen in Figure 5. It can be observed that for the three datasets containing different mode information, the neural network achieves essentially consistent recovery performance. Even for the test set not involved in training, the network can still recover relatively complete image information. This indicates that the information of the original input image has been fully mixed between high-order and low-order modes after passing through the mode mixing module. In other words, whether extracting the high-order mode image or the low-order mode image from the multimode fiber output speckle, both contain complete image information. This conclusion lays a solid foundation for using high-order modes to transmit complete image information.

### *C. Bend-Resistant Image Transmission via High-Order Modes*

To verify the bending resistance of the information carried by different modes within the multimode fiber, we collected datasets with the displacement stage positioned at 0 cm, 1 cm, 2 cm, 3 cm, and 4 cm, respectively. For the three training sets—Image A, Image B, and Image C—only the data collected at the 0 cm position were used for neural network training. After training, the test set images collected at different displacement positions were respectively input into the trained network, and the network output images were compared with the ground truth images. The calculated average Dice similarity coefficient of the test set as a function of the displacement stage position is shown in Table 1, and a comparison of the recovery results for some images is presented in Fig. 6.

**Table 1. Average DSC of the test set at different displacement stage positions.**

| Position(cm) | 0 | 1 | 2 | 3 | 4 |
|---|---|---|---|---|---|
| Image A | **0.897** | 0.798 | 0.668 | 0.527 | 0.386 |
| Image B | 0.883 | 0.623 | 0.385 | 0.281 | 0.225 |
| Image C | 0.895 | **0.886** | **0.866** | **0.826** | **0.753** |

From the data comparison in Table 1 and Fig. 6, it is evident that when only the image information from the peripheral region of the multimode fiber output speckle—i.e., the region where high-order modes predominantly contribute—is used for image information transmission, the bending stability of the system is significantly greater than that using the complete speckle information or only the central region speckle information. When only the central region speckle information is used, the images recovered by the neural network are already substantially distorted at a displacement stage position of 1 cm, and when the stage is further moved to 2 cm, the digit information becomes virtually indistinguishable, with the average DSC of the test set dropping to 0.385. In contrast, when only the peripheral region speckle information is used, the recovery performance of the model shows only a slight decline at a displacement stage position of 2 cm, and at 4 cm, the average DSC of the test set still reaches 0.753, allowing the rough shape of the digits to still be discerned. When using the complete information, a certain degree of stability can be maintained, but this is mainly attributed to the contribution of the high-order mode information.

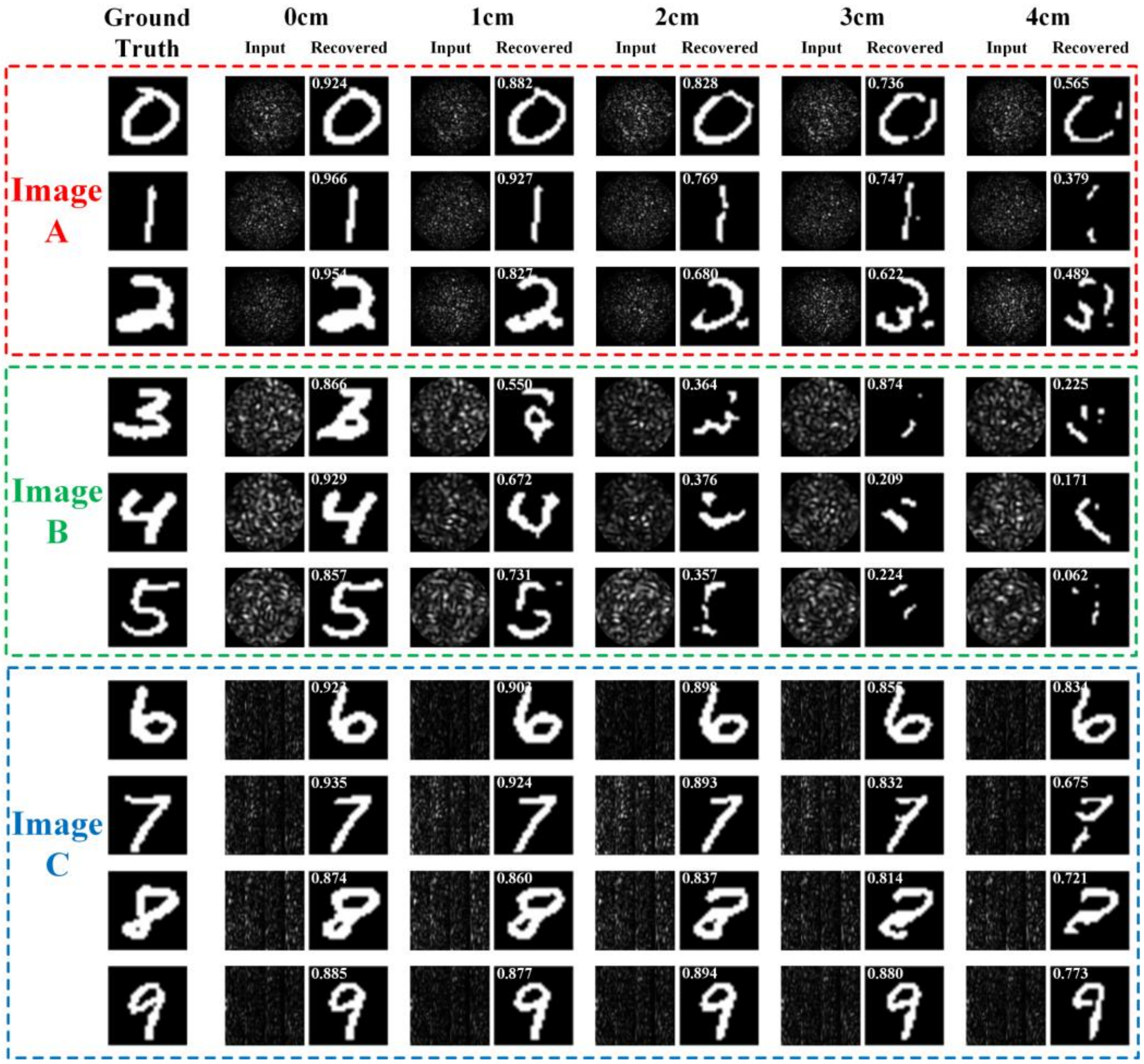

Figure 6. Comparison of recovery results for some test set images from different speckle regions when the displacement stage is at different positions.

Although the information transmission using the peripheral region of the speckle exhibits a certain degree of bending stability without any additional processing, it can be observed that the image recovery performance indeed shows a slight degradation, and there remain certain difficulties for transmitting more complex image information.

We have previously demonstrated that the image changes induced by actual bending of the multimode fiber are essentially consistent with the trends observed in theoretical simulations, indicating that the output variations are primarily governed by the bending, with other influences being virtually negligible. Furthermore, the effect of bending on all modes depends solely on the intrinsic parameters of the fiber itself. For a given fiber, regardless of the specific distribution of the input and output optical fields, the impact of bending on each identical mode contained within the optical field is exactly the same.

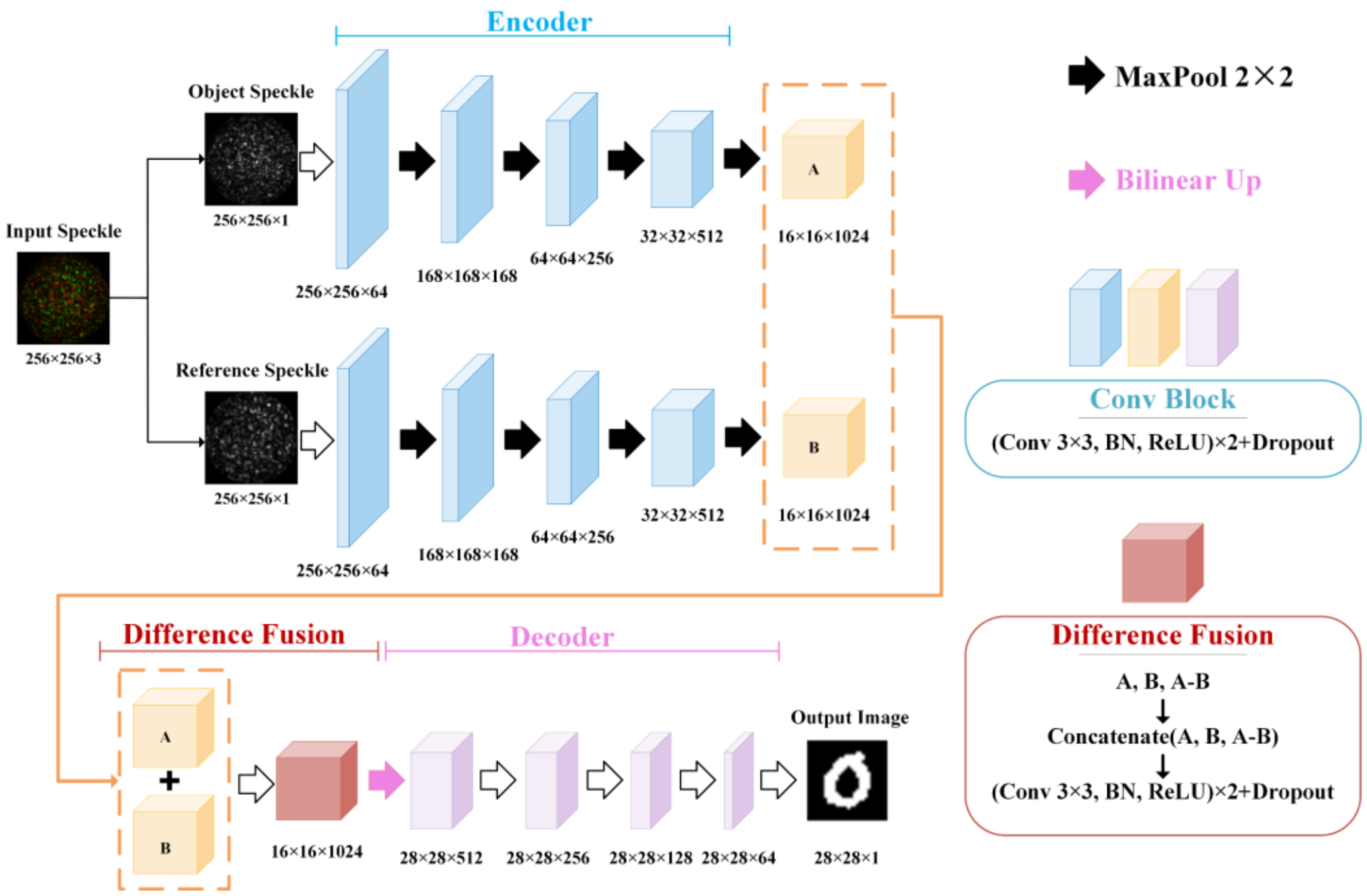

Figure 7. Architecture of the dual-encoder convolutional neural network (DE-CNN) for image reconstruction from speckle patterns using red (object speckle) and green (reference speckle) channels of an RGB input. The network takes a 256×256×3 RGB image as input, where the red and green channels contain the object and reference speckle patterns, respectively, and outputs a 28×28 reconstructed image. The two active channels are processed by two symmetric encoders, each consisting of four cascaded encoder blocks that progressively reduce the spatial resolution from 256×256 to 16×16 while increasing the number of channels from 64 to 1024. The bottleneck features A and B from the two encoders are fused via a differential module that computes A − B and concatenates it with the original features. The decoder first upsamples the fused feature maps from 16×16 to 28×28 via bilinear interpolation to avoid checkerboard artefacts, then passes through multiple convolutional layers to gradually reduce the number of channels and refine the features, and finally outputs the reconstructed image through a 1×1 convolution with sigmoid activation.

On this basis, we propose a neural network with a dual-branch encoder structure, which actively introduces a speckle image generated from a circular input field as a reference image to determine the modulation effect of the multimode fiber on the output speckle. Feature extraction is performed simultaneously on the speckle image containing object information and on the reference image acquired under the same fiber state. Both images inherently contain the same state information of the multimode fiber. By explicitly exploiting the differential features between the two through a fusion module, the target information is effectively separated from environmental perturbations, thereby enhancing the system's resistance to bending-induced disturbances. The detailed network architecture is shown in Fig. 7.

The proposed network takes an RGB image of size $256 \times 256 \times 3$ as input, where the red channel contains the object speckle pattern (the measurement carrying the object information) and the green channel contains the reference speckle pattern (the reference used for differential processing); the blue channel is unused. The two active channels are separated and fed into two identical encoder paths, denoted as Encoder A (processing the red channel) and Encoder B (processing the green channel). Each encoder consists of four cascaded blocks, with each block comprising two convolutional layers ($3 \times 3$ kernels), batch normalisation, ReLU activation, and spatial dropout, followed by a $2 \times 2$ max-pooling layer. This progressive downsampling reduces the spatial resolution from $256 \times 256$ to $16 \times 16$ while increasing the number of feature channels from 64 to 512. A bottleneck layer with two additional convolutional layers of 1024 channels further extracts high-level features from each path.

The resulting bottleneck feature maps, denoted as A and B respectively, are then fused via a differential fusion module. This module computes the element-wise difference $A - B$, concatenates this difference with the original A and B, and applies two convolutional layers to produce a fused feature map of size $16 \times 16 \times 1024$. The decoder begins by upsampling this fused feature map to $28 \times 28$ using bilinear interpolation to avoid checkerboard artefacts, then passes it through multiple convolutional layers that gradually reduce the number of channels from 512 to 64 while preserving the $28 \times 28$ resolution. Finally, a $1 \times 1$ convolutional layer with sigmoid activation outputs a single-channel reconstructed image of size $28 \times 28$.

The training settings are identical to those used for the single-encoder network described earlier: the loss function is a weighted combination of binary cross-entropy and Dice loss (equal weights), optimised with Adam (initial learning rate $1 \times 10^{-4}$, gradient clipping at 1.0). To prevent overfitting and enhance generalisation, we employ learning rate reduction (halving on plateau), early stopping (patience = 10), and model checkpointing based on validation loss.

Using the experimental setup shown in Fig. 3(a), the DMD sequentially loaded a dataset image and a circular reference image, while the CMOS camera captured the corresponding speckle images, denoted as the object speckle and the reference speckle, respectively. After each acquisition, the displacement stage was moved to an arbitrary position before the next acquisition, and a total of 1000 acquisitions were performed. The acquired images were partitioned according to the method illustrated in Fig. 3(b) and paired with the original images to form the training set, which was then fed into the neural network architecture shown in Fig. 7 for training. For better comparison of the network's optimization effect, the training set without the reference speckle was input into the previously described network without differential computation, serving as a control.

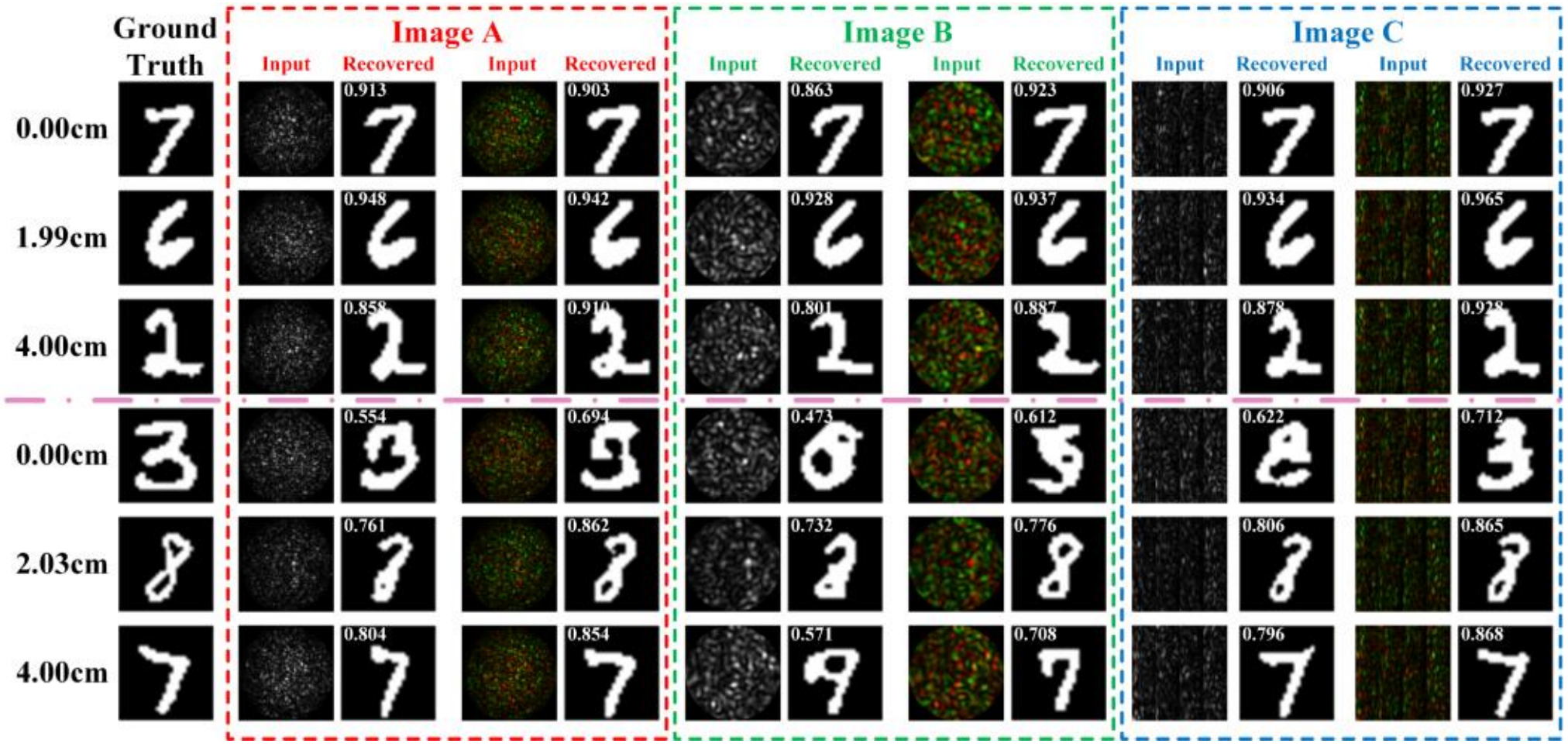

Figure 8. Comparison of recovery results for some test set images from different speckle regions when the displacement stage is at different positions. The images above the pink dashed line correspond to the 50 cm fibre length, and those below correspond to the 20 cm length.

To evaluate the impact of more severe bending conditions, we reduced the fibre length from 50 cm to 20 cm, which, with the same translation-stage travel range, decreased the minimum bending radius from 35 cm to 8 cm and thus substantially increased the bending curvature. Figure 8 presents a comparison of reconstructed images for different speckle regions and network architectures; the images above the pink dashed line correspond to the 50 cm fibre length, and those below correspond to the 20 cm length. The similarity between the reconstructed images and the ground truth was quantified using the DSC, and the average DSC values for all configurations are summarised in Table 2.

**Table 2 Average Dice similarity coefficients (DSC) under different fiber lengths and network architectures.**

| Fiber length (cm) | Network architecture | ImageA | ImageB | ImageC |
|---|---|---|---|---|
| 50cm | ED-CNN | 0.883 | 0.856 | 0.884 |
| | DE-CNN | 0.905 | 0.877 | 0.913 |
| 20cm | ED-CNN | 0.831 | 0.745 | 0.850 |
| | DE-CNN | 0.870 | 0.823 | 0.876 |

For the 50 cm fibre length, we observed no significant differences among the three image regions. Image A and Image C yielded slightly higher DSC values than Image B. The dual-encoder network (DE-CNN) improved the reconstruction performance for all three regions. When the fibre length was shortened to 20 cm, the bending curvature variation became much larger. Under the single-encoder network (ED-CNN), the reconstructions for Image A and Image C were markedly better than those for Image B. This result further indicates that the central region of the speckle pattern suffers more severe perturbations under bending. Consequently, some Image B test samples became heavily distorted and unrecognisable. Although Image A and Image C still outperformed Image B under this condition, their reconstruction quality showed a slight degradation compared with the 50 cm case. In contrast, when we applied the DE-CNN, the reconstruction for Image B improved significantly. Some fine details remained difficult to recover, but the digit information in the output images became largely identifiable. Meanwhile, the outputs for Image A and Image C also showed clear enhancement.

These results demonstrate that incorporating a reference image as a representation of the environmental state is beneficial. Using a neural network with a dual-branch encoder architecture and a differential fusion module can explicitly exploit the differences between the signal speckle and the reference speckle. This approach effectively separates the target information from environmental perturbations. As a result, the system's bending robustness is markedly improved. In addition, regardless of the network architecture, Image C consistently outperformed Image A and Image B across all test conditions. This finding further substantiates the inherent advantage of transmitting information via high-order modes. Overall, our experimental validation confirms that the proposed method, which combines high-order mode transmission through a 50/125 μm multimode fibre with a differential-fusion network, endows the multimode fibre image transmission system with excellent bending resilience.

## 4. Summary

In this paper, starting from the mode coupling theory of multimode fibers, we theoretically proved and experimentally validated that information transmission using higher-order modes in multimode fibers exhibits greater bending robustness, providing a solid theoretical foundation for experimental phenomena such as the strong correlation maintained in the peripheral region of the speckle pattern under continuous deformation.

Furthermore, we clarified that the primary cause of output image changes during multimode fiber bending is the variation in the effective propagation constants of the modes, while the energy transfer between modes plays a negligible role. This implies that the effect of bending on light propagation in a multimode fiber is independent of the modal energy distribution and depends solely on the intrinsic parameters of the fiber. In other words, for different input optical fields, although the excited mode distributions within the fiber differ, the influence of bending on the modes is identical. Based on this finding, we proposed a dual-branch differential fusion network architecture. This network takes the speckle image corresponding to a circular input field as a reference and simultaneously performs feature extraction on the speckle image containing object information and the reference image acquired under the same fiber state. Through a differential fusion module, it effectively separates the target information from environmental perturbations, further enhancing the system's resistance to bending.

It should be noted that our simulation only considered the effect of fibre bending and neglected the coupling between polarizations. Under ideal bending conditions, this approximation is valid. However, in actual fibres, core distortion and non-uniform stress distribution often induce more complex mode coupling, and the coupling between polarizations is generally not negligible. Therefore, although the simulation and experimental results exhibit consistent trends, a certain discrepancy remains.

It is worth noting that we reduced the fibre length from 50 cm to 20 cm, which significantly decreased the minimum bending radius from 35 cm to 8 cm, creating a much more severe bending environment. Under these conditions, although the single-encoder network produced severe distortions in the low-order mode region (Image B), the proposed dual-encoder differential fusion network still recovered recognisable images, and the high-order mode region (Image C) consistently maintained superior performance. These results not only validate the effectiveness of the reference-speckle-assisted differential fusion strategy, but also demonstrate that systems based on high-order mode transmission possess excellent robustness under strong bending. In summary, this work provides solid technical support for practical applications of bending-resistant imaging through multimode fibres.

**Funding.** National Key Research and Development Program of China (2021YFF0307804).

**Disclosures.** The authors declare no conflicts of interest.

**Data availability.** Data underlying the results presented in this paper are not publicly available at this time but may be obtained from the authors upon reasonable request.

## Reference

1. I. N. Papadopoulos, S. Farahi, C. Moser and D. Psaltis, "High-resolution, lensless endoscope based on digital scanning through a multimode optical fiber," Biomedical Optics Express 260-270,(2013/02/01)
2. H. Cao, T. Čižmár, S. Turtaev, T. Tyc and S. Rotter, "Controlling light propagation in multimode fibers for imaging, spectroscopy, and beyond," Advances in Optics and Photonics 524-612,(2023/06/30)
3. S. M. Popoff, G. Lerosey, R. Carminati, M. Fink, A. C. Boccara and S. Gigan, "Measuring the Transmission Matrix in Optics: An Approach to the Study and Control of Light Propagation in Disordered Media," Physical Review Letters 100601,(03/08/)
4. S. Popoff, G. Lerosey, M. Fink, A. C. Boccara and S. Gigan, "Image transmission through an opaque material," Nature Communications 81,(2010/09/21)
5. M. Plöschner, T. Tyc and T. Čižmár, "Seeing through chaos in multimode fibres," Nature Photonics 529-535,(2015/08/01)
6. M. Cui and C. Yang, "Implementation of a digital optical phase conjugation system and its application to study the robustness of turbidity suppression by phase conjugation," Optics Express 3444-3455,(2010/02/15)
7. D. B. Conkey, A. M. Caravaca-Aguirre and R. Piestun, "High-speed scattering medium characterization with application to focusing light through turbid media," Optics Express 1733-1740,(2012/01/16)

8. L. V. Amitonova and J. F. de Boer, "Compressive imaging through a multimode fiber," Optics Letters 5427-5430,(2018/11/01)
9. Z. Wang, H. Li, T. Zhong, Q. Zhao, V. R. V, H. Li, Z. Yu, J. Pu, Z. Chen, X. Yuan and P. Lai, "Speckle-driven single-shot orbital angular momentum recognition with ultra-low sampling density," Nature Communications 11097,(2025/12/12)
10. L. Huang, Q. Li, C. Zhang, H. Gao, W. Li, X. Xu, J. Qi, X. Liu, Z. Wen and Q. Yang, "Deep Learning for Disturbance-Resistant Compressive Sensing Multimode Fiber Imaging," Laser & Photonics Reviews e02398,(2026/05/01)
11. B. Rahmani, I. Oguz, U. Tegin, J.-l. Hsieh, D. Psaltis and C. Moser, "Learning to image and compute with multimode optical fibers," Nanophotonics 1071-1082,(2022/02/01)
12. Z. Wen, Z. Dong, Q. Deng, C. Pang, C. F. Kaminski, X. Xu, H. Yan, L. Wang, S. Liu, J. Tang, W. Chen, X. Liu and Q. Yang, "Single multimode fibre for in vivo light-field-encoded endoscopic imaging," Nature Photonics 679-687,(2023/08/01)
13. S. Li, C. Saunders, D. J. Lum, J. Murray-Bruce, V. K. Goyal, T. Čižmár and D. B. Phillips, "Compressively sampling the optical transmission matrix of a multimode fibre," Light: Science & Applications 88,(2021/04/21)
14. S. Yan, Y. Sun, F. Ni, Z. Liu, H. Liu and X. Chen, "Image reconstruction through a nonlinear scattering medium via deep learning," Photonics Research 2047-2055,(2024/09/01)
15. P. Fan, Y. Wang, M. Ruddlesden, X. Wang, M. A. Thaha, J. Sun, C. Zuo and L. Su, "Deep Learning Enabled Scalable Calibration of a Dynamically Deformed Multimode Fiber," Advanced Photonics Research 2100304,(2022/10/01)
16. Z. Liu, L. Wang, Y. Meng, T. He, S. He, Y. Yang, L. Wang, J. Tian, D. Li, P. Yan, M. Gong, Q. Liu and Q. Xiao, "All-fiber high-speed image detection enabled by deep learning," Nature Communications 1433,(2022/03/17)
17. Z. Wu, R. Gao, S. Zhou, F. Wang, Z. Li, H. Chang, D. Guo, X. Xin, Q. Zhang, F. Tian and Q. Wu, "Robust Super-Resolution Image Transmission Based on a Ring Core Fiber with Orbital Angular Momentum," Laser & Photonics Reviews 2300624,(2024/02/01)
18. Z. Wu, R. Gao, J. Zhu, F. Wang, H. Chang, Z. Li, D. Guo, L. Zhu, Q. Zhang, X. Huang, J. Yan, L. Jiang and X. Xin, "Dynamic perturbation mitigation via polarization difference neural network for high-fidelity ring core fiber image transmission," Optics Express 33305-33320,(2024/09/09)
19. J. Wang, H. Zhang, S. Zeng, H. Wen, J. Wang and X. Zhang, "High-fidelity image transmission and reconstruction through multimode fiber using OAM modes and deep learning," Optics Express 4917-4932,(2026/02/09)
20. D. Loterie, D. Psaltis and C. Moser, "Bend translation in multimode fiber imaging," Optics Express 6263-6273,(2017/03/20)
21. S. Li, S. A. R. Horsley, T. Tyc, T. Čižmár and D. B. Phillips, "Memory effect assisted imaging through multimode optical fibres," Nature Communications 3751,(2021/06/18)

22. R. Xu, L. Zhang, B. Xu, Z. Qian and D. Zhang, "Anti-perturbation multimode fiber speckle imaging and recognition through learning invariant fiber characteristics hidden in speckle patterns," Optics & Laser Technology 112961,(2025/10/01/)
23. F. T. S. Yu, J. Zhang, S. Yin and P. B. Ruffin, "Analysis of a fiber specklegram sensor by using coupled-mode theory," Applied Optics 3018-3023,(1995/06/01)
24. H. Taylor, "Bending effects in optical fibers," Journal of Lightwave Technology 617-628,(
25. L. Palmieri and A. Galtarossa, "Coupling Effects Among Degenerate Modes in Multimode Optical Fibers," IEEE Photonics Journal 1-8,(
26. H. G. Unger and E. Conwell, "Planar Optical Waveguides and Fibres," Physics Today 51-54,(
27. R. Ulrich, S. C. Rashleigh and W. Eickhoff, "Bending-induced birefringence in single-mode fibers," Optics Letters 273-275,(1980/06/01)
28. R. T. Schermer and J. H. Cole, "Improved Bend Loss Formula Verified for Optical Fiber by Simulation and Experiment," IEEE Journal of Quantum Electronics 899-909,(
29. R. T. Schermer, "Mode scalability in bent optical fibers," Optics Express 15674-15701,(2007/11/26)